\documentclass[12pt,preprint]{aastex}

\def\lesssim{\mathrel{\hbox{\rlap{\hbox{\lower4pt\hbox{$\sim$}}}\hbox{$<$}}}}
\def\gtrsim{\mathrel{\hbox{\rlap{\hbox{\lower4pt\hbox{$\sim$}}}\hbox{$>$}}}}

\usepackage{graphicx}
\newcommand{\dF}{{^{^*}\!\!F}}

\newcommand{\bF}{{\bf F}}
\newcommand{\bU}{{\bf U}}

\newcommand{\del}{{\partial}}
\newcommand{\bsq}{{{||b||^2}}}
\newcommand{\CTB}{{{\mathcal{B}}}}
\newcommand{\lum}{{{\mathcal{L}}}}
\newcommand{\prim}{{{\mathbf{P}}}}
\newcommand{\sI}{{\mathcal{I}}}

\newcommand{\sJ}{{\mathcal{J}}}
\newcommand{\beq}[1]{\begin{equation} #1 \end{equation}}

\newcommand{\deriv}[2]{\frac{ d #1 }{ d #2 }}
\newcommand{\pderiv}[2]{\frac{ \partial #1 }{ \partial #2 }}

\newcommand{\GMT}{{2003ApJ...589..444G}}
\newcommand{\dVH}{{2003ApJ...589..458D}}
\newcommand{\dVHK}{{2003ApJ...599.1238D}}
\newcommand{\NT}{{1973blho.conf..343N}}

\begin{document}

\title{DIRECT CALCULATION OF THE RADIATIVE EFFICIENCY OF AN ACCRETION DISK AROUND A BLACK HOLE}

\author{Scott C. Noble, Julian H. Krolik}
\affil{Physics and Astronomy Department\\
Johns Hopkins University\\ 
Baltimore, MD 21218}
\and
\author{John F. Hawley}
\affil{Astronomy Department\\
University of Virginia\\ 
P.O. Box 400325\\
Charlottesville, VA 22904-4325}

\email{scn@jhu.edu; jhk@jhu.edu; jh8h@virginia.edu}

\begin{abstract}

Numerical simulation of  magnetohydrodynamic (MHD) turbulence makes it
possible to study accretion dynamics in detail.  However, special effort
is required to connect inflow dynamics (dependent largely on angular
momentum transport) to radiation (dependent largely on thermodynamics
and photon diffusion).  To this end we extend the
flux-conservative, general relativistic MHD code \texttt{HARM} from
axisymmetry to full 3D.  The use of an energy conserving algorithm
allows the energy dissipated in the course of relativistic accretion
to be captured as heat.  The inclusion of a simple optically thin
cooling function permits explicit control of the simulated disk's
geometric thickness as well as a direct calculation of both the
amplitude and location of the radiative
cooling associated with the accretion stresses.  
Fully relativistic ray-tracing is used to compute the
luminosity received by distant observers.  For a disk with aspect ratio
$H/r \simeq 0.1$ accreting onto a black hole with spin parameter $a/M =
0.9$, we find that there is significant dissipation beyond that predicted
by the classical Novikov-Thorne model.  However, much of it occurs deep
in the potential, where photon capture and gravitational redshifting can
strongly limit the net photon energy escaping to infinity.  In addition,
with these parameters and this radiation model, significant thermal and
magnetic energy remains with the gas and is accreted by the black hole.
In our model, the net luminosity reaching infinity is $6\%$ 
greater than the Novikov-Thorne prediction.  If the accreted
thermal energy were wholly radiated, the total luminosity of the
accretion flow would be $\simeq 20\%$ greater than the Novikov-Thorne value.

\end{abstract}


\section{Introduction}

    For the past thirty-five years, it has been the standard view in the
astrophysical community that the total amount of energy per unit mass
dissipated in the course of accretion onto a black hole is exactly equal
to the binding energy of the innermost stable circular orbit \citep{\NT};
consequently, it depends only on the black hole spin parameter $a/M$.  The
argument leading to this result depended on a number of assumptions: that
the flow is time-steady and axisymmetric, that any heat dissipated is
promptly radiated, and that the $r$-$\phi$ component of stress goes
to zero at the ISCO.  Although the first several assumptions
appear to be relatively innocuous, the last was regarded as questionable
almost from the beginning \citep{T74} and has been subject to renewed
questioning in more recent years \citep{K99,G99}.  If accretion flows
were fundamentally hydrodynamic, the heuristic argument for this boundary
condition (that the inertia of matter in the plunging region should
always be much smaller than the inertia in the stable-orbit portion of
the disk) would be cogent; the point raised by all its critics is that
if magnetic fields play an important role, their stress would not necessarily
be diminished even in regions of small matter density.  Resolution of
this point is important because continued forces at and inside the ISCO
would permit continued dissipation, possibly substantially increasing
the total.

     Although the significance of magnetic forces was
no more than a speculation when
the zero-stress boundary condition was first criticized, in recent
years it has been recognized that they are, in fact, essential to
accretion \citep{BH98}.  Stimulated by this recognition, the past
decade has seen many numerical simulations of global disk dynamics
incorporating magnetic forces under the assumption of ideal
magnetodhyrodynamics (MHD)
\citep{HK01,HK02,AR01a,AR01b,AR03,MM03,\dVHK,KHH05,GSM04}.  Initially
these simulations assumed Newtonian dynamics in a pseudo-Newtonian
potential; in the middle of this effort, new codes were developed
that permit simulations in 
general relativity \citep{\dVH,\GMT}.  So far, while often treating
angular momentum transport quite accurately, all of these simulations
have handled thermodynamics and energy transport comparatively crudely:
In {\tt GRMHD}, the code developed by De Villiers and Hawley, only an
internal energy equation is solved, in which the gas is assumed
to behave adiabatically except in shocks; in this code, therefore,
there are sizable (and uncontrolled) numerical losses of energy
whenever magnetic field or kinetic energy is lost on the gridscale.
By contrast, in \texttt{HARM}, the code developed by Gammie et~al., a total
energy equation is solved (so no energy is lost), but there are
also no radiative losses.  The best effort that could be made to
estimate actual radiative efficiency was therefore through plausible,
but {\it ad hoc} models, usually keyed to the magnetic stress
\citep{2008arXiv0801.2974B}.

     In an effort to remedy this situation, we have altered the
\texttt{HARM} code in two significant ways.  First, we have extended it from
2D (axisymmetric) to 3D.  This extension has two major consequences:
we can study nonaxisymmetric fluctuations, and are free from 2D
artifacts like the ``channel solution"; and we are not limited by
the anti-dynamo theorem to short duration simulations.  Second, we
have introduced a toy-model optically thin cooling function.  By
this means, we can track how much radiation might be produced (and
where) in order to compute the radiative efficiency explicitly.  We
can also use this cooling function to regulate the geometric thickness
of the accretion flow.  A detailed description of the new code
(which we call \texttt{HARM3D}) can be found in \S~2.

       For our first use of this code, we chose to run a simulation
that would illustrate how MHD turbulence influences the global energetics
of accretion onto a black hole.  Its results can be compared directly
to those of NT: Time-averages of its data can be matched against the
classical model's steady state.  Quantities integrated over constant-radius
shells can be compared with the corresponding vertically-integrated ones
derived assuming axisymmetric and ``razor-thin" disks.  The cooling function
can be designed to (almost) reproduce the prompt radiation assumption.
However, we have no need
to impose any guessed boundary condition on the stress because in this
numerical calculation we are able to use the the real physical boundary
condition to accretion dynamics: the black hole's event horizon.  Thus, the
ratio of the energy radiated in this simulation to the mass accreted in it
provides a direct test of how much the zero-stress boundary condition
affects the radiative efficiency.  In addition, of course, we will also
be able to examine the interesting effects of non-stationary flow,
non-axisymmetry, and so on.

      Because we recognize that quantitative results may well depend
on a number of parameters (magnetic field configuration and disk thickness,
most notably) and because our radiation model does not fully represent
any particular physical situation, we emphasize that the numbers we
present here are only preliminary samples.  When we discuss these
results, we will explain more specifically the degree to which they
are model-dependent.  We intend to explore more
fully in future work both how to model this process more realistically
and how external parameters such as
magnetic configuration and accretion rate couple with black hole
spin to control the radiative output of accretion onto black holes.

\section{The Computation: \texttt{HARM3D} and the Parameters of Our Simulation}
\label{sec:theory-and-methods}

 
Quite a number of general relativistic MHD simulation codes have been
written already
\citep{1999MNRAS.303..343K, 1999ApJ...522..727K, 2003ApJ...589..444G,
2003ApJ...589..458D, 2005PhRvD..72b4028D, 2005PhRvD..72d4014S,
2005ApJ...635..723A, 2006ApJ...637..296A, 2006ApJ...641..626N,
2006astro.ph..9004M, 2006CQGra..23.6503A, 2007MNRAS.379..469T,
2007ApJ...668..417F, 2007A&A...473...11D, 2008arXiv0804.4572C}. 
Our starting point for the code used in this paper was the \texttt{HARM} code
\citep{2003ApJ...589..444G,2006ApJ...641..626N}.  \texttt{HARM}
solves the equations of motion in flux-conservative form, but
is restricted to axisymmetry\footnote{Technically, our 
code's procedural flow and data structure design developed from 
an early 3D version of the \texttt{HAM} code \citep{gammie-comm}
that is now publicly available as a shearing box 
code \citep{codelib,2008ApJS..174..145G}.  All other 
routines were either developed by us, or taken from the 
public version of \texttt{HARM} found in \cite{codelib}.}.  
As we have already mentioned, axisymmetric
calculations suffer from two major drawbacks: 
the dominance of ``channel solutions", which are ubiquitous
in 2D but unstable in 3D \citep{BH98}, and the fact that neither turbulence
nor magnetic field can be sustained indefinitely in 2D.
To avoid these limitations, we
extended the algorithm to three spatial dimensions.  
\texttt{HARM}'s conservative
formulation means that it does not lose energy to numerical
dissipation; rather, kinetic and magnetic energies lost at the
gridscale are captured as heat.   At the same time,
a conservative
formulation permits easy introduction of a formal radiative cooling term.
Thus, our new code, called \texttt{HARM3D}, is the first global MHD accretion
code to treat thermodynamics in a controlled fashion\footnote{Concurrent 
with our effort, \cite{2008arXiv0808.2860S} have also developed a 3D
version of the HARM code.  They have now used this to explore
the dynamical effects of cooling in a disk around a Schwarzschild
black hole, but did not directly measure the radiation it produced.}.

\subsection{Basic Equations}
\label{sec:basic-equations}

We begin the description of \texttt{HARM3D} with an explicit
statement of the equations
governing our model.  Contrasts with \cite{2003ApJ...589..444G}  (\texttt{HARM})
and  \cite{2003ApJ...589..458D} (\texttt{GRMHD}) will be highlighted along the way.
We use Greek letters for spacetime indices, and 
Roman letters for spacelike indices.  The signature of the metric is the same
as the one used in \cite{MTW} (i.e., -+++), and geometrized units are used such
that  $G=c=M=1$.

The general relativistic MHD (GRMHD) equations of motion include the continuity equation, 
\beq{
\nabla_\mu \left( \rho u^\mu \right) = 0 \quad , \label{continuity-eq}
}
the equations of local energy conservation
\beq{
\nabla_\mu {T^{\mu}}_\nu  = 0 \quad , \label{energy-conservation-eq}
}
and Maxwell's equations
\beq{
\nabla_\nu \dF^{\mu \nu}  = 0 \quad ,  \label{maxwell-eq-1}
}
\beq{
\nabla_\nu F^{\mu \nu}  = J^\mu \quad .   \label{maxwell-eq-2}
}
Here, $\rho$ is the rest-mass density, $u^\mu$ is the 4-velocity of the 
fluid, $F^{\mu \nu}$ is the Faraday tensor times $1/\sqrt{4\pi}$,
$\dF^{\mu \nu}$ is the dual of this tensor or the Maxwell tensor times 
$1/\sqrt{4\pi}$, and $J^\mu$ is the 4-current\footnote{We follow \cite{\GMT}
in our definition of the electromagnetic field tensor and magnetic field variables.}.
The total stress-energy tensor is the sum of the fluid part, 
\beq{
T^{\mu\nu}_\mathrm{fluid} = \rho h u^\mu u^\nu 
+ P g^{\mu\nu}  , \label{fluid-stress}
}
and the electromagnetic part 
\beq{
T^{\mu\nu}_\mathrm{EM} = F^{\mu \lambda} {F^\nu}_\lambda 
- \frac{1}{4} g^{\mu \nu} F^{\lambda \kappa}  F_{\lambda \kappa} 
= \bsq u^\mu u^\nu + \frac{1}{2} \bsq g^{\mu\nu} - b^\mu b^\nu \quad 
 , \label{em-stress}
}
where $g_{\mu \nu}$  is the metric, $h =\left(1 + \epsilon + P/\rho \right)$ is
the specific enthalpy, $P$ is the pressure, $\epsilon$ is the 
specific internal energy density, $b^\mu = \dF^{\nu \mu} u_\nu$ 
is the magnetic field 4-vector, and $\bsq \equiv b^\mu b_\mu$ is 
twice the magnetic pressure $P_m$\footnote{The magnetic 4-vector $b^\mu$
defined in this paper is equivalent to that in \texttt{HARM} and \texttt{GRMHD},
even though our and \texttt{HARM}'s definition is different from \texttt{GRMHD}'s
by a sign.  For this reason, \texttt{GRMHD}'s version of our 
equation~(\ref{maxwell-eq-2}) differs by a sign.  These sign
differences can all be reconciled by noting that their electromagnetic field 
tensors have opposite sign.
The resulting equations of motion are independent of these sign conventions.}.

Equations~(\ref{continuity-eq}-\ref{maxwell-eq-1}) can be expressed 
in flux conservative form 
\beq{
\del_t \bU\left(\prim\right) = 
-\del_i \bF^i\left(\prim\right) + \mathbf{S}\left(\prim\right) \, 
\label{conservative-eq}
}
where $\bU$ is a vector of ``conserved'' variables, $\bF^i$ are the fluxes, 
and $\mathbf{S}$ is a vector of source terms.  Explicitly, these 
are 
\beq{
\bU\left(\prim\right) = \sqrt{-g} \left[ \rho u^t , {T^t}_t 
+ \rho u^t , {T^t}_j , B^k  \right]^T
\label{cons-U}
}
\beq{
\bF^i\left(\prim\right) = \sqrt{-g} \left[ \rho u^i , {T^i}_t + \rho u^i , {T^i}_j , 
\left(b^i u^k - b^k u^i\right) \right]^T
\label{cons-flux}
}
\beq{
\mathbf{S}\left(\prim\right) = \sqrt{-g} 
\left[ 0 , {T^\kappa}_\lambda {\Gamma^\lambda}_{t \kappa} 
, {T^\kappa}_\lambda {\Gamma^\lambda}_{j \kappa} , 0 \right]^T \, 
\label{cons-source}
}
where $g$ is the determinant of the metric, ${\Gamma^\lambda}_{\mu \kappa}$ is
the metric's affine connection, and $B^i = \dF^{it}$ is our magnetic field 
\footnote{The ``CT field'' of \texttt{GRMHD}, $\CTB^i$, is proportional
to our magnetic field:  $\CTB^i = \sqrt{-4 \pi g} B^i$}.
Note that the source term for the energy equation is non-zero only
when the metric is time-dependent (as evidenced by its proportionality
to $\Gamma^{\lambda}_{t\kappa}$).
The equations of motion are closed by an equation of state,
$P = \left(\Gamma - 1 \right) \rho \epsilon$, 
where $\Gamma$ is the adiabatic index, set to $5/3$ in this work. 
The primitive variables, $\prim=\{\rho, P, \tilde{u}^i\}$, are recovered 
using an optimized version of the the ``2D'' algorithm described in
\cite{2006ApJ...641..626N}.  The primitive velocity is the flow's velocity as 
viewed by a zero angular momentum observer (ZAMO):
\beq{
\tilde{u}^i =  u^i + \alpha W g^{ti} \quad , \label{primitive-velocity}
}
where $\alpha = 1/\sqrt{-g^{tt}}$ is the lapse function and
$W=\alpha u^t$ is the Lorentz factor.

\subsection{Initial Data}
\label{sec:ic_bc}

In the initial state of the simulation, the matter is in an axisymmetric
hydrostatic torus that orbits the black hole with a specific angular
momentum profile
slightly shallower than Keplerian and $u^r = u^\theta = 0$.
The disk is centered about the equator of the black hole's spin and is initially
assumed to be isentropic.  In curved spacetimes, the angular
frequency---$\Omega = u^\phi/u^t$---is not a simple function of the
specific angular momentum---$l = -u_\phi/u_t$.  For example, one can
show that when $u^r = u^\theta = 0$ in Boyer-Lindquist coordinates,
$\Omega = \left( g^{t\phi} - g^{\phi\phi} l \right)/
                 \left(g^{tt} - g^{t\phi} l \right)$.
In order to solve the time-independent Euler equations, we must therefore
specify $l(r,\theta)$.  Following \cite{chak85} and \cite{2003ApJ...599.1238D},
we do this by assuming that
$\Omega \sim \lambda^{-q}$, where $\lambda^2 = l/\Omega$.
The solution is simplified by setting $\lambda$ to its Schwarzschild value
$\lambda = \sqrt{-g^{tt}/g^{\phi\phi}}$, which is exact when $a=0$ but leads to a
solution marginally out of equilibrium when $a \neq 0$;  the slight departure
from equilibrium insignificantly affects the disk's evolution because the
magnetic field quickly becomes dynamically important.  Ultimately,
we arrive at an equation for $l(r,\theta)$:
$l/l_\mathrm{in} = \left(\lambda/\lambda_\mathrm{in}\right)^{2-q}$,  where
$l_\mathrm{in} = l(r_\mathrm{in},\pi/2) $ and
$\lambda_\mathrm{in} = \lambda(r_\mathrm{in},\pi/2)$.

With the intention of closely mimicking the initial
conditions of simulation KDP of \cite{2003ApJ...599.1238D}, we 
put the torus pressure maximum at $r=25M$ and choose an angular
momentum distribution parameter $q=1.67$.  The torus inner boundary is
$r_\mathrm{in} = 15M$, with $l_\mathrm{in} = 4.576$.
These parameters yield a disk very similar to that of 
De~Villiers et~al., but with a slightly larger $l_\mathrm{in}$.

The solution to Euler's equations provides us with $h$ and $u^\mu$.
The rest-mass density is then calculated from
the equations of state---$P = \left(\Gamma - 1 \right) \rho \epsilon$ and
$P = K \rho^\Gamma$---and $h$:
$\rho = \left[\left(h-1\right) \left(\Gamma - 1\right) / 
      \left(K\Gamma\right) \right]^{\left(1/(\Gamma-1)\right)}$.
We suppose that the gas is non-relativistic, choosing $\Gamma = 5/3$
and $K=0.01$.  Integrating over the volume of the initial gas distribution,
we find a total rest-mass of 353.  This is $20\%$ larger than
that in simulation KDP, a shift due to our slightly different choice
of $l_\mathrm{in}$.  Note that the code units of gas mass are completely
arbitrary.

The initial magnetic field lies entirely within the torus and follows contours of
constant density.  The magnitude of the magnetic field is set so that
the volume-weighted integrated magnetic pressure is $100$ times less
than the volume-weighted integrated gas pressure.  

The atmosphere surrounding the disk is unmagnetized and static.  The
atmosphere's density and pressure are set to their smallest allowed values,
which are chosen so that the floor state is in approximate pressure equilibrium:
$\rho_\mathrm{floor} = 7\times10^{-9}\rho_\mathrm{max} r^{-3/2}$ and
$P_\mathrm{floor} = 7\times10^{-11}\rho_\mathrm{max} r^{-5/2}\left(\Gamma-1\right)$,
where $\rho_\mathrm{max}$ is the initial maximum value of the
rest-mass density in the disk.

\subsection{Radiative Cooling}
\label{sec:radiative-cooling}

A magnetized accretion disk is subject to the magneto-rotational
instability (MRI), which transfers angular momentum outward.  This transfer
taps into the available free energy of differential rotation, creating
the magnetic fields and poloidal velocity fluctuations that make
up the resulting MHD turbulence.  This turbulence is dissipative;
magnetic and kinetic energy is lost numerically at the gridscale.
Equation~(\ref{conservative-eq}), however, ensures that in the numerical
solution all that dissipated energy is converted to heat.  If that heat
were retained by the fluid, the disk would become ever hotter and
geometrically thicker.  Ultimately the thermal energy would either be
accreted by the hole or be carried out from the disk by a wind.
By adding a loss term to the energy equation, we can estimate either
the luminosity of those systems in which radiation is efficient or
the total heat generated in those systems in which it is not.
We assume that the radiation described by this loss term
is optically thin.  It therefore acts as a passive sink in the local 
energy conservation equation~(\ref{energy-conservation-eq}): 
\beq{
\nabla_\mu {T^{\mu}}_\nu  = -\mathcal{F}_\nu \quad , 
\label{energy-conservation-eq-with-cooling}
}
where $\mathcal{F}_\nu$ is the amount of 
radiated energy-momentum per unit 4-volume in the coordinate frame.  To
describe the radiation, we make the simplest assumption: that the emission 
is isotropic in the fluid's frame:
\beq{
\mathcal{F}_\nu = \lum u_\nu \label{radiative-flux}
}
where the ``cooling function'' $\lum$ is the rate energy is radiated per unit
proper time in the fluid frame.

The NT assumptions include complete prompt radiation of all
locally-dissipated heat.  We cannot exactly replicate that in a
simulation, for the gas must retain some thermal energy.  However,
we can arrange for the great majority of the heat to be radiated by
constructing a cooling function that keeps the temperature of the gas at
a small fraction of the virial temperature.  In so doing, we can also
control the disk's aspect ratio $H/r$, a parameter often considered
significant in analytic disk models.

In different contexts, different definitions of the scale-height $H$
are sometimes used.  For a thin isothermal disk in a Newtonian potential,
the density profile is Gaussian, $\rho \propto \exp[-z^2/(2H_G^2)]$, with
$H_G^2 = c_i/\Omega$, for isothermal sound speed $c_i$.  Another common
measure of the scale-height is the half-width at half-maximum (HWHM),
$H_{HWHM} = \sqrt{2\ln 2}H_G$.  A third is the vertical density moment,
\begin{equation}
H \equiv \int \, d\theta d\phi \sqrt{-g} \, \rho \sqrt{g_{\theta\theta}}
     |\theta - \pi/2| \, / \int \, d\theta d\phi \, \sqrt{-g} \rho .
\end{equation}
When the profile is Gaussian, $H = \sqrt{2/\pi}H_G = 0.798 H_G$.


We prefer the moment definition because it is a direct measure of the
characteristic mass-weighted disk thickness, it is
robust with respect to fluctuations, and it is closely related to the
characteristic scalelengths of hydrostatic balance.  Ideally, it
would be computed in the fluid-frame, but in the interest of computational
economy we define it in the coordinate frame.  Moreover, when any of
these definitions of disk thickness is taken in ratio to the radius, it
should be recalled that the radial coordinate $r$ is not a proper distance.  
Unfortunately, there is no obvious adequate substitute. 

In any event, given this definition, the temperature that should produce a
desired aspect ratio $H/r$ in Newtonian gravity is
\beq{
T_*(r) = \frac{\pi}{2}\left[ \frac{H}{r} r \Omega(r) \right]^2  \quad .
\label{target-temperature}
}
In code units, $T_* = (\Gamma-1)\epsilon = (2/3)\epsilon$.

In our simulation, we evaluate $T_*$ in the disk body using the relativistic
orbital frequency $\Omega(r > r_\mathrm{isco}) = 1/\left(r^{3/2} + a/M\right)$.
In a more completely relativistic
treatment, $\Omega(r)$ would be replaced by $\Omega_K R_z^{1/2}(r)$, where
$\Omega_K$ is the Newtonian Keplerian rotation frequency and $R_z$ is the
relativistic correction factor for the vertical gravity (\citep{alp97};
notation as in \citep{k99book})\footnote{The expression given in these
references contains a typo: $E_{\infty}$ should be $E_{\infty}^2$.  In
addition, we define $R_z$ inside the ISCO by setting $u_\phi$ and $u_t$
to the values they have at the ISCO.}.
Inside the innermost stable circular orbit (ISCO), we define
$\Omega$ as the orbital frequency of a particle with the specific
energy and angular momentum of the circular orbit at the ISCO:
\beq{
\Omega(r < r_\mathrm{isco}) = 
\frac{ g^{\phi \mu}\left(r,\theta=\pi/2\right) \   K_\mu }{
g^{t \mu}\left(r,\theta=\pi/2\right)  K_\mu}.
\label{omega-within-isco}
}
Here $K_\mu$ is the 4-velocity of the ISCO orbit. 

To ensure that the disk stays near the target temperature,
the cooling rate should be rapid (i.e., $\sim \Omega$), but drop to zero
when the temperature falls below $T_*(r)$.  All of these
criteria are satisfied by a cooling function with the form
\beq{
\lum = s \Omega \rho \epsilon \left[ Y - 1 + \left| Y - 1 \right| \right]^q 
\quad , 
\label{cooling-function}
}
where $Y = (\Gamma-1)\epsilon / T_*(r)$ and $s$ is a constant of proportionality.
Note that the term in the square brackets serves as a switch, so that
$\lum=0$ whenever $Y < 1$.  The exponent $q$ controls how rapidly the
cooling rate grows when the temperature exceeds the target.  We found that
$q=1/2$ cools the plasma efficiently while maintaining
a stable evolution, and we set $s=1$.   Only those fluid elements
on bound orbits---where $\left( 1 + \epsilon + P/\rho \right) u_t > -1 $---are 
cooled.

In addition to controlling the vertical thickness of the disk,
the cooling function provides a self-consistent way of comparing
emission from the simulated disk with that expected in a standard NT
model.  When making this comparison, we use 
the angle-averaged fluid-frame luminosity per unit area (of an annulus located
at the equator) from our 3D simulation data: 
\beq{
F_\mathrm{ff}(r)  =  \frac{ \int \int dx^{\left(\phi\right)} dx^{\left(\theta\right)} \lum }{
\int dx^{\left(\phi\right)} |_{\theta=\pi/2}}   \quad , 
\label{fluid-frame-flux}
}
where each component of the vector 
$dx^{\left(\mu\right)} = {e^{\left(\mu\right)}}_{\nu} \, dx^\nu$ 
represents the extent of a cell's dimension as measured in the fluid element's
rest frame, and ${e^{\left(\mu\right)}}_{\nu}$ is the orthonormal tetrad that
transforms vectors in the Boyer-Lindquist coordinate frame to the local fluid
frame (see \cite{2008arXiv0801.2974B} for explicit expressions for the tetrad).  
The vector $dx^\nu$ is the Boyer-Lindquist coordinate frame version of the
Kerr-Schild vector $dx^\nu_\mathrm{KS} = 
\left[0,\Delta r, \Delta \theta, \Delta \phi\right](r,\theta,\phi)$, 
where $\Delta r$, $\Delta \theta$, $\Delta \phi$ are the radial, poloidal and
azimuthal extents of our simulation's finite volume cell located at
$(r,\theta,\phi)$. 

We also wish to calculate the radiated luminosity  measured by a distant
observer in order to include the effect of photon losses into the black hole. 
This is done by ray-tracing through the spacetime and integrating the 
radiative transfer equation along geodesics.  Redshift factors from differences
in inertial reference frames are automatically taken into account 
and include such effects as gravitational redshift and relativistic beaming.
As with the cooling function, we assume that the fluid is optically thin
and---consequently---ignore scattering and absorption.   

 To briefly summarize our method, we trace a large number of rays from
observers at infinity at 8 polar angles and integrate the transfer
equation along each ray.  Since we aim only at estimating the bolometric
luminosity of the disk, and not at computing its spectrum, we assume that
all radiated energy is emitted at a single frequency equal to the Doppler
shifted frequency of observation.  From the transfer solution along each ray,
we construct images of the disk
as it would be seen by each of those observers, and then sum the radiation they
receive.   In order to compute the radiation reaching infinity for the
NT model (whose photons are also subject to possible capture
by the black hole and Doppler shifting), we place an emissivity designed to
match the NT surface brightness in the two planes of cells nearest
the equatorial plane.  Assuming that the four-velocities
of those cells are exactly those of circular orbits at those radii, we
then compute the luminosity at infinity in this model by the same ray-tracing
technique as employed on our simulation data.  Additional details are given in
Appendix~\ref{sec:radi-transf-calc}.  

\subsection{Coordinates, grid, and boundary conditions}

Equation~(\ref{conservative-eq}) is solved using finite volume techniques on a
uniform grid in the so-called ``Modified Kerr-Schild'' (MKS) coordinate system 
described in \cite{\GMT}.  It is based on the Kerr-Schild (KS) coordinate 
system that eliminates the coordinate singularity at the horizon.  
The modification allows us to adjust the radial and angular discretization
through a continuous coordinate transformation.  We set the MKS parameter
$h_{MKS}=0.3$ \citep{\GMT}, which makes the poloidal cell scale at the axis
about $5.7$ times larger than that at the 
equator and allows us to resolve greater detail in the accretion 
disk than would be possible with the same number of equally
spaced grid cells.

The simulation reported here used $192\times192\times64$ cells in the radial, 
poloidal, and azimuthal directions respectively, with 
$r \in [1.28M,120M]$, $\theta \in [0.05\pi,0.95\pi]$, $\phi \in [0,\pi/2]$.
The radial extent is as large as the one used in KDP except our coordinates
penetrate the horizon by five cells.  
We tested whether our polar angle discretization adequately
resolved the fastest-growing mode of the magneto-rotational instability by 
calculating---in the local fluid frame---the 
fastest-growing mode's wavelength and the local poloidal size of a cell. 
Averaged over azimuth and time (over $t=[7000M,15000M]$),
the fastest-growing mode was resolved by at least seven cells at 
all radii.  The absolute minimum number of cells
per wavelength for all time and radii is never smaller than four. 
For the time discretization, we have found a Courant factor of $0.8$ 
is adequate when used with the existing step size control method in \texttt{HARM}.

In the original MKS coordinate system, 
cells are placed all the way to the axis.  We have introduced a new 
reflecting boundary condition that allows us to excise the 
coordinate singularity there.  With the boundary placed at an angle of
$0.05 \pi$ from the axis (as in KDP), the excision enlarges the
time step we can take, speeding up the evolution by about a factor of 
four relative to simulations using grids without the cut-out.  The radial
boundary conditions are the same as in the released version of \texttt{HARM}, 
and we use periodic boundary conditions for the azimuthal boundaries. 

\subsection{Algorithmic details}
The equations of motion are integrated using almost the same high-resolution
shock-capturing methods as described in \cite{2003ApJ...589..444G}.  We still 
use \texttt{HARM}'s Lax-Friedrichs numerical flux formula, 
as it is more diffusive than the HLL formula and seems to be stabler
for our purposes.  However, the piecewise-linear reconstruction
is replaced with a parabolic 
interpolation method \citep{1984JCoPh..54..174C} 
as our means of reconstructing values at cell faces.  As in
\texttt{HARM}, we use an MC (monotonized central-differenced) slope limiter.
Parabolic reconstruction improves stability in low density 
regions where $\bsq/\rho >> 1$, such as are found in the funnel 
\citep{2006Mckinney}.  

We also use parabolic interpolation in the ``Flux-CT'' scheme of 
\cite{2000JCoPh.161..605T}
that preserves the divergence constraint.  Originally, the
electromotive forces (EMFs)
at the cell faces were calculated by a  
second-order accurate two-point averaging procedure.  This method failed 
to dissipate a cell-scale sawtooth instability seen in the 
magnetic field  along the intersection between the inner radial and 
poloidal boundaries.  
Parabolic interpolation of the EMFs, however, is successful 
at quelling the instability.

Even with the improvements described so far, stably 
evolving plasma whose total energy is dominated by magnetic and
kinetic energies is difficult.  
In a conservative code like \texttt{HARM3D}, the critical step is deriving good
primitive variables, $\prim$, from the conserved quantites.
For instance, the magnetic field is typically a few orders of magnitude 
larger than the pressure in the funnel.  The pressure is recovered
from inverting the equation for the total energy, ${T^t}_t$, which involves
subtracting ${T_\mathrm{EM}^t}_t$ and the fluid's inertia term from ${T^t}_t$.
In the funnel, this operation is essentially a subtraction of two large
numbers whose result will likely be the size of either term's truncation error. 
The subtraction can result in either positive or negative pressures. 
This is known as the ``positive-pressure problem'' in hydrodynamics and MHD and 
has been studied extensively \citep{1993ApJ...414....1R,1999JCoPh.148..133B}.
We have found that even when positive pressures are recovered,  numerical errors
may result in pressure fluctuations that differ by orders of 
magnitude between adjacent cells.  These fluctuations create pressure
gradients that can accelerate matter and add energy to the system artificially. 

In order to treat the positive pressure problem and correct for 
other unphysical states that may arise (e.g. $W<0$), we have completely
redesigned \texttt{HARM}'s recovery procedure.  The most significant change 
is the inclusion of the conservation of  entropy equation
\beq{
\nabla_\mu \left( \mathcal{S} u^\mu \right) = 0 \label{entropy-eq}
}
where 
\beq{
\mathcal{S} \equiv \frac{P}{\rho^{\Gamma-1}} \quad . \label{entropy-density}
}
Following a method similar to that of \cite{1999JCoPh.148..133B}, we integrate 
equation~(\ref{entropy-eq}) in parallel with (\ref{conservative-eq}).  Whenever
the standard primitive variable method fails to converge, $\tilde{u}^i$ is
unphysical, or 
$\rho \epsilon < 10^{-2} \bsq$, we use a new inversion method 
which is identical to the standard one except the total energy equation is 
replaced by equation~(\ref{entropy-density}).  Even though this inversion method 
is guaranteed to yield a positive pressure, it can either fail to converge
to a solution or yield an unphysical $\tilde{u}^i$.  If either happens, 
we interpolate $\prim$ using data from neighboring cells for which we have
successfully
calculated $\prim$.  Finally, we impose a floor on the pressure and density and 
ensure $W \le 50$ by renormalizing $\tilde{u}^i$.   

We note that using equation~(\ref{entropy-eq})
leads to a method that no longer conserves total energy to round-off error,
but the impact of these departures from strict conservation is limited.
The entropy equation is substituted for the energy equation only where
the fluid is very strongly magnetically-dominated, and only when no
energy-conserving method yields a physical solution.  In the simulation
reported here, the net injection or loss of mass,
energy and angular momentum is only $\sim 0.001 - 0.007$ times the flux of 
these quantities through the numerical domain.  After the period of initial
transients, the places where non-conservative effects can be found are
almost exclusively restricted to the edge of the axial cut-out and the region
roughly $45^{\circ}$ from the polar axis within the ergosphere.

 
We have verified that our new code is second-order accurate for smooth
solutions and satisfactorily passes the tests described in \cite{2003ApJ...589..444G}.
A  quantitative comparison of our code's performance to that of \texttt{GRMHD}
will be left for future work.

\section{Results}

      Our initial condition is a torus of gas in hydrostatic equilibrium,
entirely contained within the simulated volume; our goal is to present results
characteristic of an accretion flow with a fixed aspect ratio in a long-term
equilibrium.  Before quoting results directly from the simulation data, we must
therefore do two things: demonstrate that the fixed aspect ratio is achieved,
and define more precisely the degree to which the simulation is in a
statistical steady-state with respect to inflow.

\subsection{Scale-height regulation}

\begin{figure}
\epsscale{0.5}
\plotone{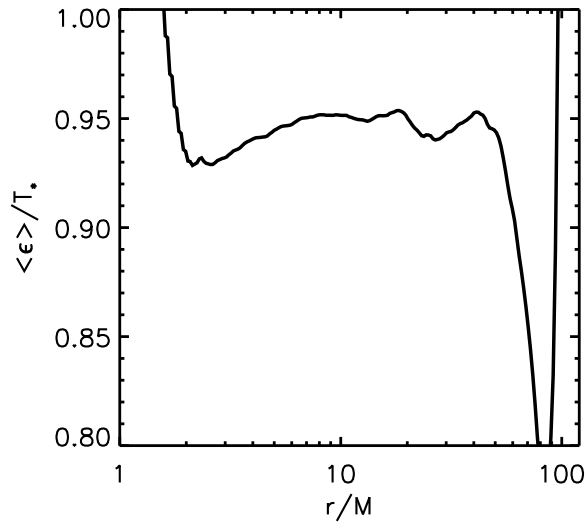}
\caption{Ratio of mean temperature $(\Gamma-1)\langle \epsilon \rangle$ to
target temperature $T_*$. The time-averaging interval was 7000--$15000M$.
\label{fig:tempcontrol}}
\end{figure}

      We set the parameters of our cooling function so that the ratio of the
sound speed to the local orbital speed would produce a disk with a constant
aspect ratio $H/r = 0.13$.  In Figure~\ref{fig:tempcontrol}, we show how
well the temperature was held to $T_*$ by comparing the time-averaged
volume-weighted temperature in the bound accretion flow to the local value
of $T_*$.  In the main disk body, this mean value was about (0.93--$0.95)T_*$,
but it rises sharply inside the ISCO.  In other words, our cooling function
succeeded in holding the disk temperature very close to (in fact, slightly
below) the target temperature, but inside the ISCO, where the inflow time
becomes comparable to or shorter than the cooling time, the temperature
rises well above $T_*$.

\begin{figure}
\epsscale{0.5}
\plotone{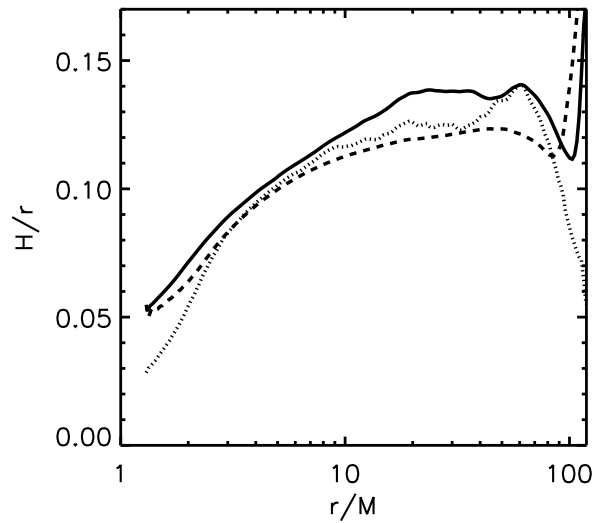}
\caption{Time-averaged density scale-height as a function of radius (solid
curve), and time-averaged HWHM (dotted curve).  The data were sampled every
$20M$ from $t=7000M$ to $15000M$.
Hydrostatic scale-height assuming the shell- and time-averaged temperature
but employing the relativistic correction described in
\S~\ref{sec:radiative-cooling} (dashed curve).
\label{fig:scaleheight}}
\end{figure}

How well our temperature-regulation led to a disk aspect ratio matching the
goal value of 0.13 can be seen in
Figure~\ref{fig:scaleheight}.  The actual $H/r$
was slightly above the goal ($\simeq 0.14$) through most of the simulation
volume, but with a tendency to diminish inward inside $r=20M$.  At $r=10M$,
$H/r \simeq 0.12$; by the time the flow reaches the ISCO, it is only
$\simeq 0.07$.  Comparison with the curve showing how the scale-height
changes as a result of including the relativistic correction to the vertical
gravity (as discussed in \S~\ref{sec:radiative-cooling}) demonstrates
that this thinning at small radius can be largely attributed
to neglect of that effect.
Thus, use of our cooling function achieved its principal goal: to place the
scale-height of the disk under explicit control.

\begin{figure}
\epsscale{0.5}
\plotone{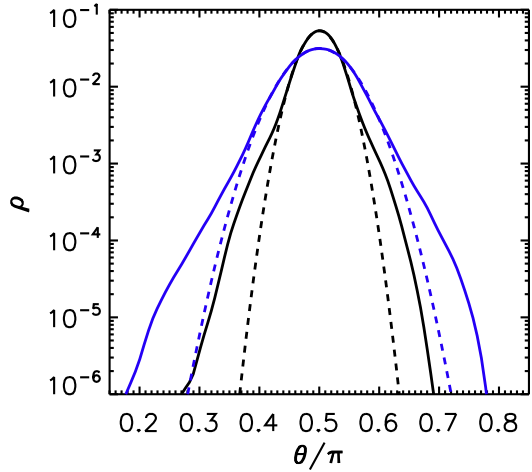}
\caption{Time- and azimuthally-averaged density (solid curves) at
the ISCO (black) and $r=12M$ (blue), each fit to a Gaussian (dashed curves).
\label{fig:Gaussian}}
\end{figure}

    Because our cooling function has a target temperature
depending only on radius, at any particular radius the gas in the
main body of the disk is nearly isothermal, and the density profile is
therefore close to Gaussian (Fig.~\ref{fig:Gaussian}).  At higher
altitudes above the midplane, the density falls slower than
the Gaussian, presumably due to magnetic support.  For this reason,
the moment scale-height is slightly greater than the HWHM
(Fig.~\ref{fig:scaleheight}).

     We chose a value of $H/r$ small enough that a key approximation of
the NT theory could be approximately replicated in the simulation: the prompt
radiation of dissipated heat.  However, if the disk is to have a finite thickness,
it cannot radiate all its heat.  The parameters we chose for the cooling
function yielded an accretion rate-weighted mean specific enthalpy that was
well-described by $h \simeq 1 + 0.031 (r/M)^{-0.8}$.  At large radius,
where the Newtonian approximation applies, the ratio of $h-1$ to the net
binding energy is $\simeq 0.06 (r/M)^{0.2}$, while at the ISCO this ratio is
$\simeq 0.1$.  Thus, this toy-model does assure that the majority of the
dissipated heat is radiated.

\subsection{Inflow equilibrium}

\begin{figure}
\epsscale{1}
\plottwo{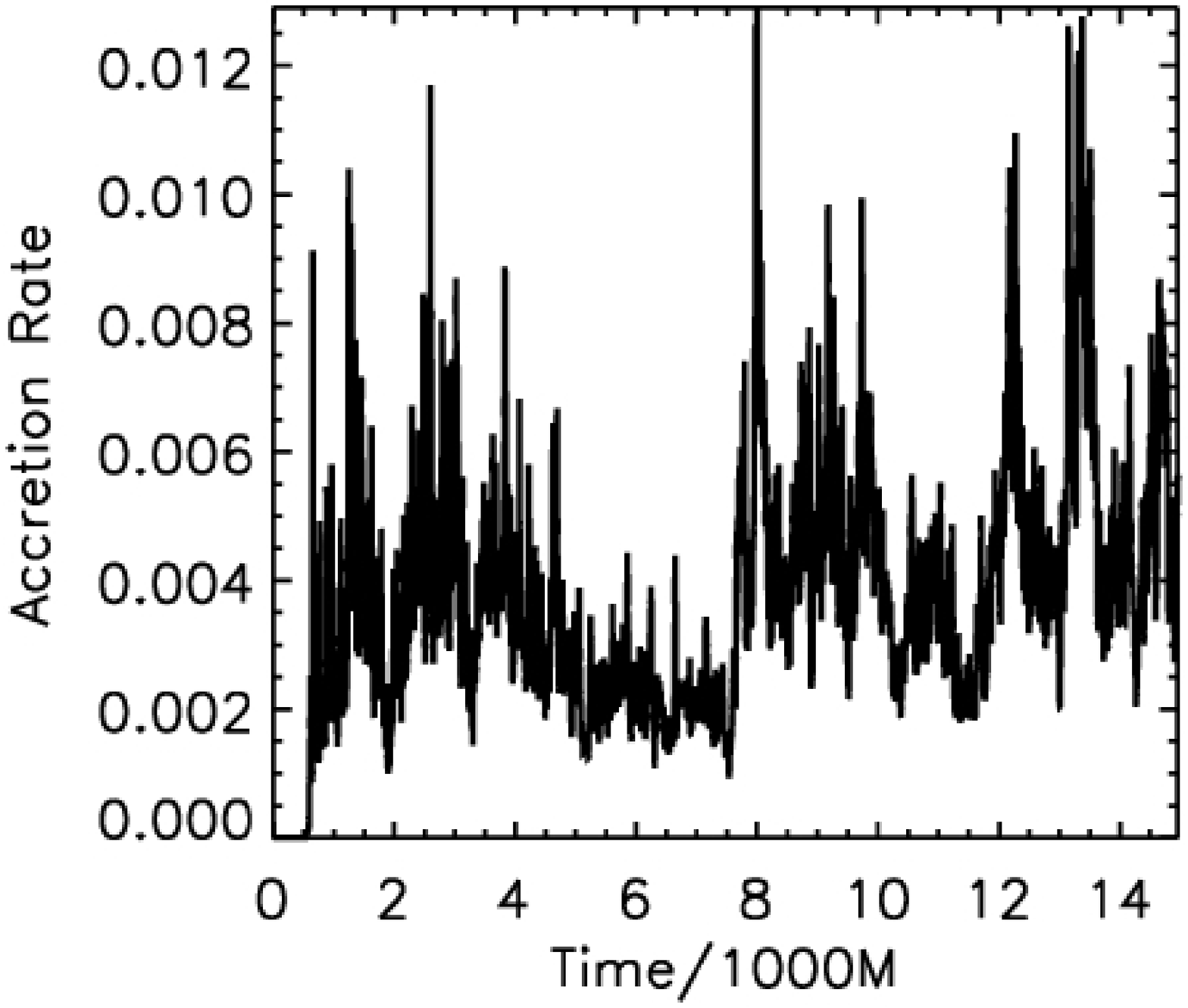}{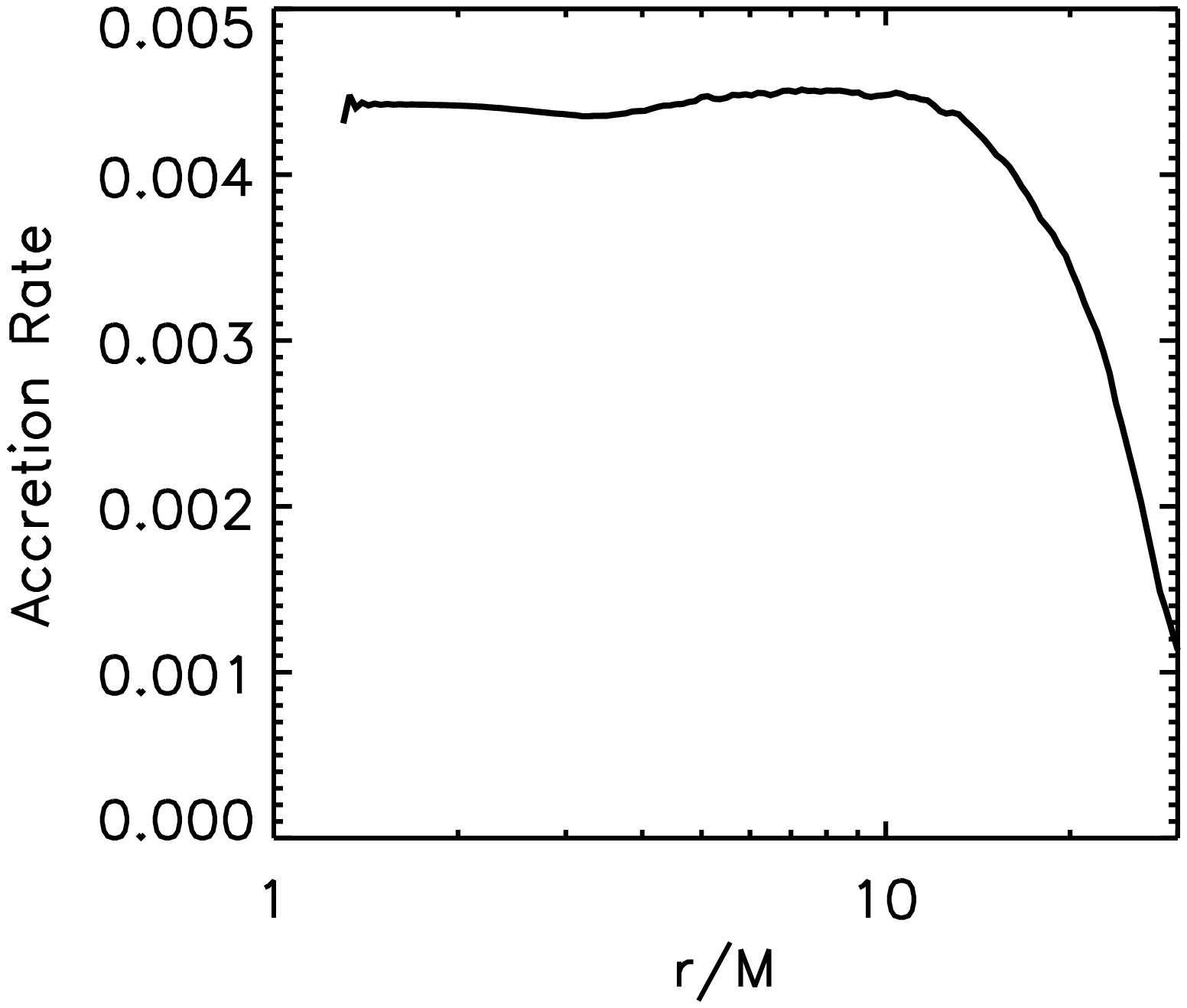}
\caption{(left) The accretion rate (in code units) at the event horizon
as a function of time.  A rate of 0.005 translates to accreting a
fraction 0.14 of the initial mass in a time of $10000M$.  (right) The
shell-integrated accretion rate as a function of radius,
averaged from $t=7000M$ to $15000M$, sampled every $1M$.
\label{fig:inflowequil}}
\end{figure}

\begin{figure}
\epsscale{0.5}
\plotone{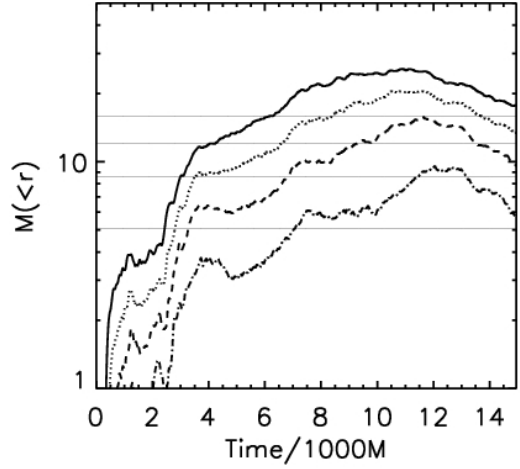}
\caption{The mass contained within four sample radii: $14M$ (solid curve),
$12M$ (dotted curve), $10M$ (dashed curve), and $8M$ (dash-dot
curve), all as functions of time.  The thin solid lines mark $90\%$ of
the final mass for each of these radii.  A mass of 10 in code units
is $2.8\%$ of the initial torus mass.
\label{fig:massfillin}}
\end{figure}

     If the accretion flow were in a strict steady-state, the local
(i.e., shell-integrated) mass accretion rate $\dot M(r)$ would be the
same at all radii at all times and the mass interior to a given radius
would likewise be constant.  In these turbulent disks fed by a finite
mass reservoir, the most we can hope for is that the time-average local
accretion rate is nearly constant as a function of $r$ through most of
the accreting region, and the mass of the inner disk, after an
initial period of growth, eventually levels off and fluctuates within
some range.  The degree to which we approach these goals is shown in
Figures~\ref{fig:inflowequil} and \ref{fig:massfillin}.  In the left-hand
panel of Figure~\ref{fig:inflowequil}, we see that the accretion rate
(measured at the event horizon) varies by roughly a factor of five in an
extremely irregular way.  Nonetheless, as shown in the right-hand panel,
the time-averaged $\dot M(r)$ is very nearly constant from the horizon
to $r\simeq 14M$ for the latter $8000M$ of the simulation.  The reason
why we choose the interval $7000M$--$15000M$ for averaging is shown in
Figure~\ref{fig:massfillin}.   As this figure demonstrates, it takes
roughly the first $7000M$ of the simulation for the mass of the inner
disk to reach a rough plateau.  Because the mass interior to
a given radius fluctuates, we chose the starting point for time-averaged
quantities to be the point at which essentially all the inner disk
had reached at least $90\%$ of its final mass, which is approximately
$t = 7000M$.

However, for the purposes of estimating the radiative efficiency, we
require a tighter definition of inflow equilibrium.   This is because
we wish to contrast the computed radiation rate with the NT
rate at an accuracy of a few percent or better.  In the NT model,
$23\%$ of the total light is emitted between $12M$ and $25M$, 
where our simulation shows significant departures from
inflow equilibrium; a further $27\%$ comes from outside $25M$, where
our simulation is not an accretion flow and we do not compute
the luminosity at all.  For these reasons, when we contrast the NT
luminosity with that produced in the simulation, we adjust the local
accretion rates to mimic inflow equilibrium and attach a carefully-chosen
representation of large-radius emission where needed (see below for details).

\subsection{Explicit radiative efficiency}

\begin{figure}
\epsscale{1}
\plottwo{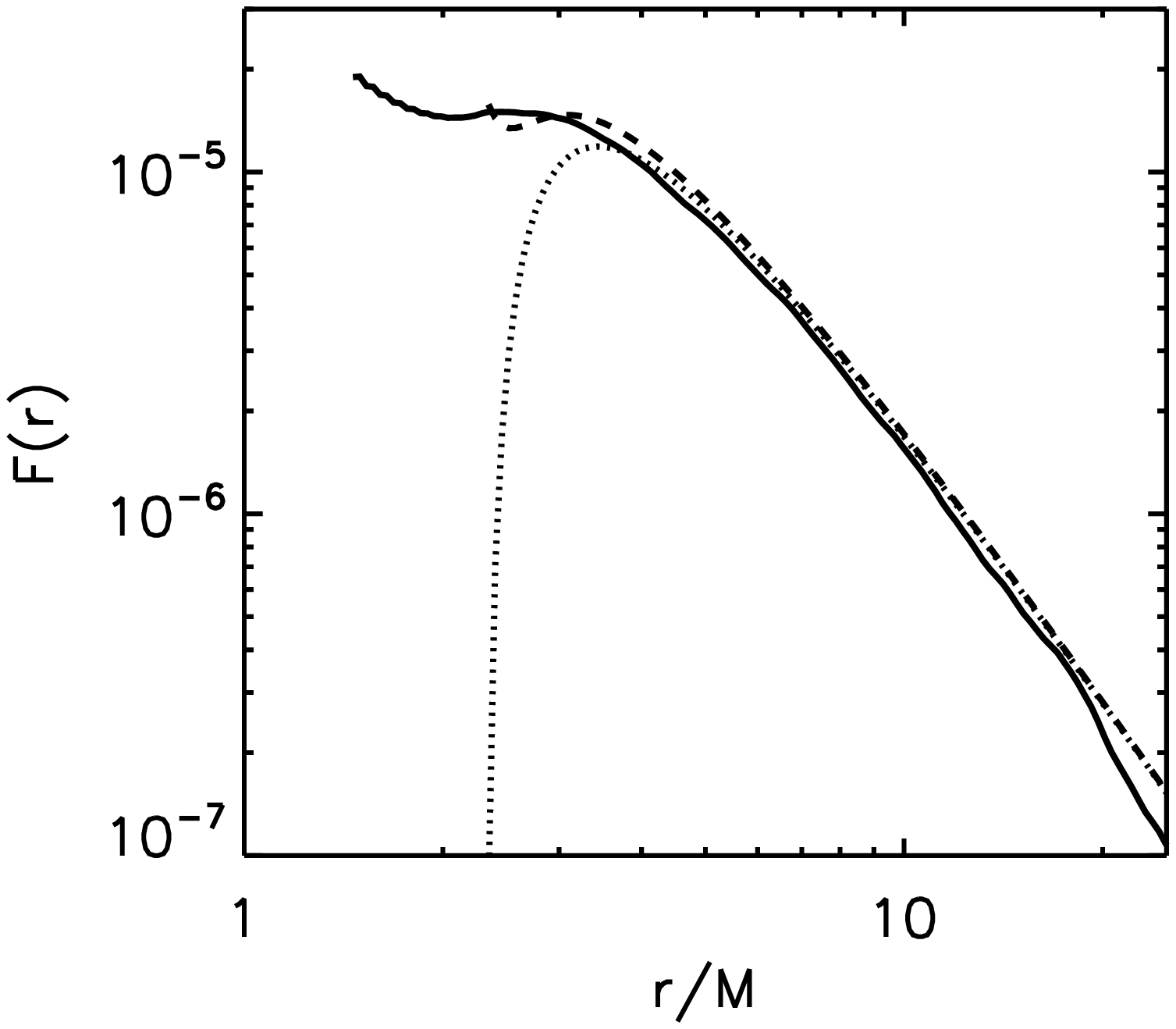}{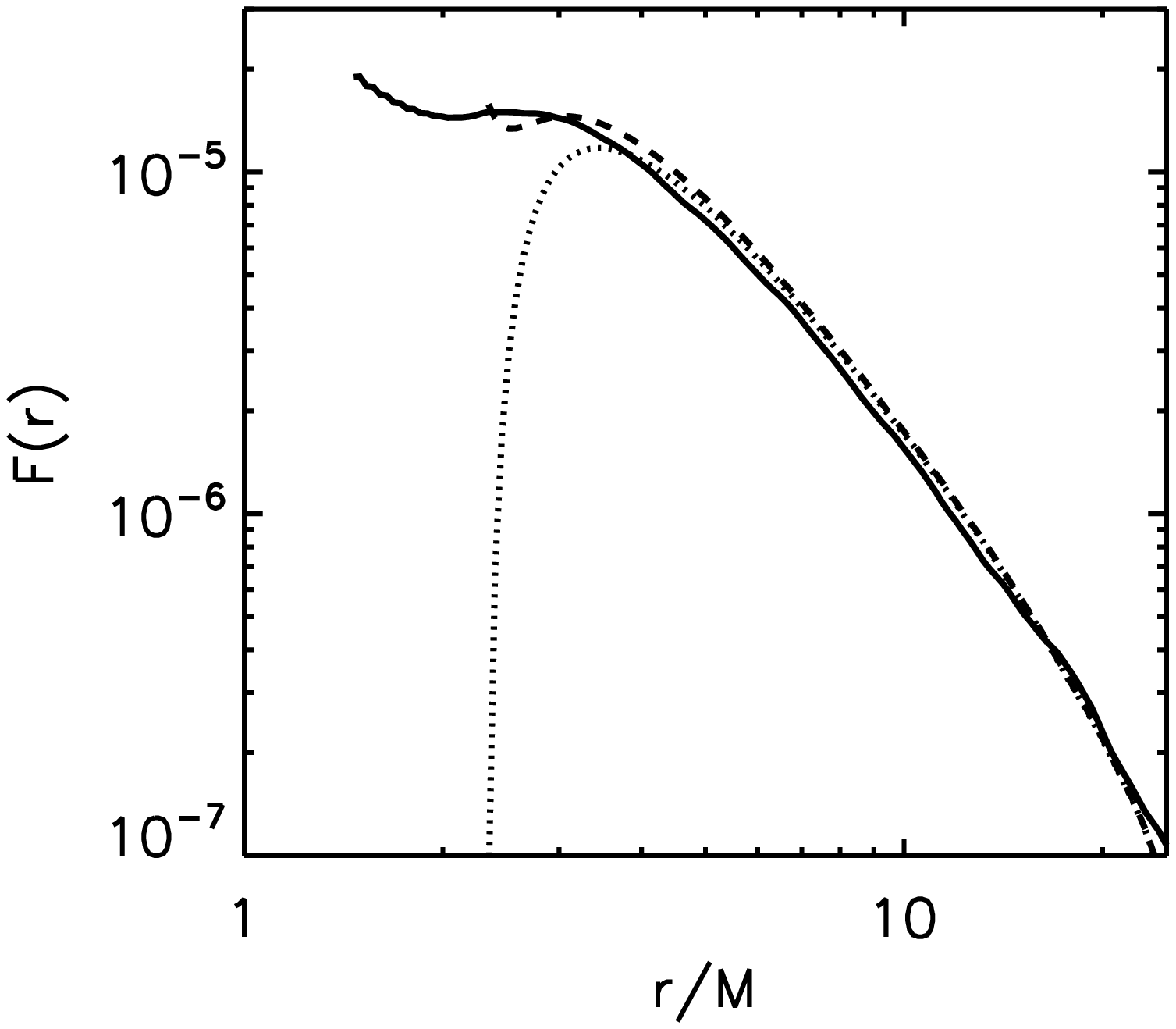}
\caption{Radiated flux per unit area in the fluid frame as a function of radius:
time-averaged simulation data (solid curve); as predicted by the
NT model (dotted curve); as predicted by the AK model with
$\Delta \epsilon = 0.01$ (dashed curve). (left) Using the time-averaged data from
$7000M$ to $15000M$.  (right) Adjusting the NT and AK emissivities as
described in the text.
\label{fig:surfbright}}
\end{figure}

The fluid-frame emissivity $F_{ff}(r)$ found in the simulation and the NT
prediction for this quantity are displayed in Figure~\ref{fig:surfbright}.  In the
leftmost panel, we show how they compare when the NT emissivity
uses the time-averaged accretion rate at the horizon over the same interval
for which the simulation data were averaged, i.e., $t=7000M$--$15000M$.
For precise comparison of the two radiation models, we
remove the effects of deviations from inflow equilibrium by altering
the NT emissivity so that the value at any given radius 
corresponds to the time-averaged accretion rate at that radius,
as determined by the simulation.  Put another way, the fluid-frame surface
brightness in a truly time-steady NT model may be written as
$(3/4\pi)(GM\dot M/r^3)R_R(r)$ (notation as in \cite{k99book}); we adjust
this to $(3/4\pi)(GM\dot M(r)/r^3)R_R(r)$, with $\dot M(r)$ the time-averaged
accretion rate at radius $r$ in the simulation.   By doing so, we compare
the two radiation models in a way that factors out any contrasts due solely to
fluctuations in the accretion rate.  The adjusted version is shown in
the right-hand panel of Figure~\ref{fig:surfbright}.

As this pair of figures shows, the two models coincide closely in the main disk
body, but contrast sharply near and within the ISCO, which is at
$r\approx 2.3M$ for this spin.  Because the NT model is founded on
energy and angular momentum conservation in a time-steady disk, this
coincidence is no surprise where the influence of the NT no-stress
boundary condition is small.  The principal departure between the two,
a systematic offset in which the simulation curve lies $\simeq 10\%$ below the
NT curve, is due to the small fraction of the dissipated heat that the
gas must retain to provide vertical pressure support in the
disk.  Near the ISCO, the accretion-weighted mean
specific enthalpy is $\simeq 0.018$ greater than unity.  This thermal
energy is $12\%$ of the binding energy at the ISCO (0.155 per unit rest-mass).

At small radius, however, the fluid-frame surface brightness of the
simulation differs substantially from the NT model.  At $r=3M$,
the simulation surface brightness is greater by $40\%$; at the ISCO,
although the NT model would predict no radiation, the
surface brightness indicated by the simulation is roughly the same
as at $r=3M$; close to the horizon, the surface brightness rises
to about twice the maximum predicted by the classical model.

Another way to characterize the contrast between the simulation results
and the NT model is through the intermediary of another analytic model.
In the model of \cite{ak00}, it is supposed that a finite
stress is exerted at the ISCO, but all other assumptions follow those
of NT.  This model (which we will abbreviate as AK) is parameterized
by the additional efficiency $\Delta \epsilon$ due to the non-zero
stress at the ISCO; in the curves shown in the two panels of
Figure~\ref{fig:surfbright}, $\Delta \epsilon = 0.01$, a value chosen
as an approximate best fit between the AK model and the simulation
data.  In the region
immediately outside the ISCO, where the AK model is defined, it does
a reasonable job of reproducing the simulation results, particularly
when allowance is made for the retained heat.

\begin{figure}
\epsscale{1}
\plottwo{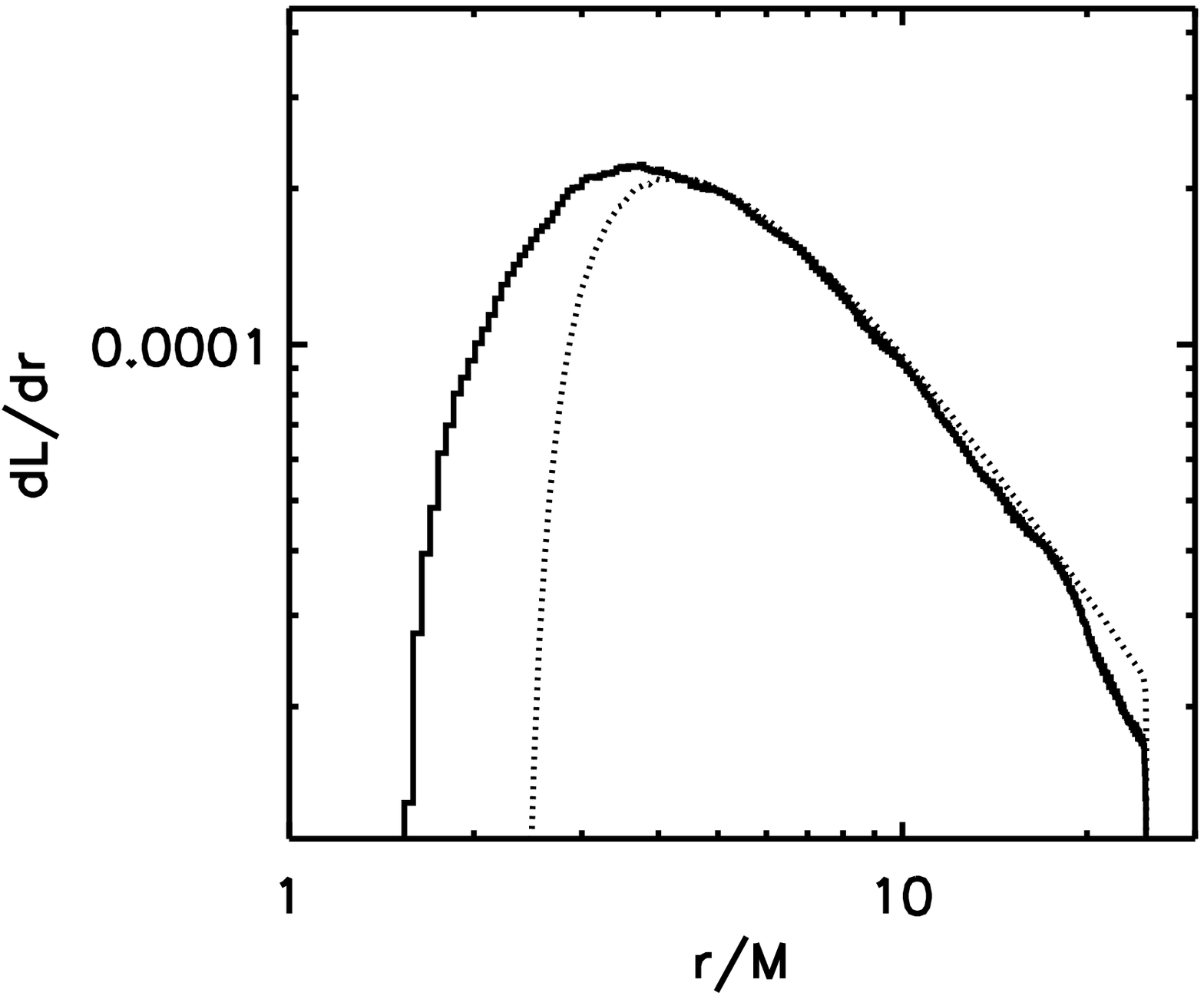}{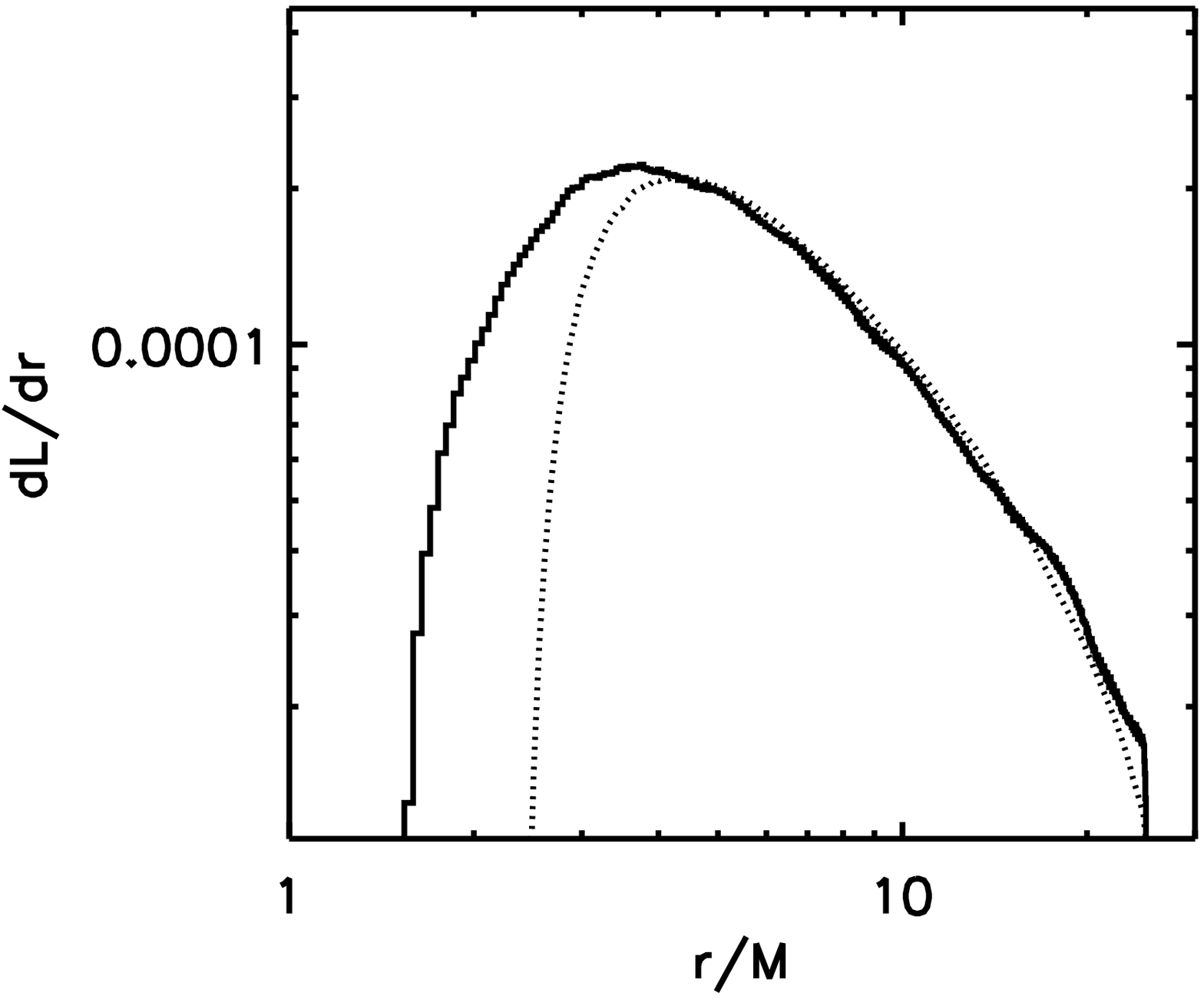}
\caption{Luminosity received at infinity per unit radial coordinate:
time-averaged simulation data (solid curve); as predicted by the
NT model (dotted curve). (left) Using the time-averaged data from
$7000M$ to $15000M$.  (right) Adjusting the NT emissivity as
described in the text.
\label{fig:dldr}}
\end{figure}

Only some of this radiation reaches infinity, and any that does
arrives with a significant Doppler shift, most often toward the red.
Using the techniques described in \S~\ref{sec:radiative-cooling} and
Appendix~A, we computed the luminosity received at infinity per unit radial
coordinate $dL/dr$, which is
shown in Figure~\ref{fig:dldr}.   Like the emissivity in
the fluid frame, $dL/dr$ for the simulation data in the main disk body
closely tracks the NT prediction.  The only difference between the two
is that the simulation data version lies slightly ($\simeq 10\%$) below
the NT curve: this offset is simply another reflection of the offset
already seen in the fluid-frame emissivity due to the non-zero heat
content of a physical disk.  At small radii, the shelf in
the fluid-frame emissivity is transformed into
an inward extension of significant luminosity that extends from $r \simeq 4M$
to $r \simeq 2M$.  Although the fluid-frame emissivity extends
farther inward, its efficiency in creating luminosity at infinity is
cut off by a combination of increasing redshift and probability
of photon capture by the black hole.

At larger radii, departures from inflow equilibrium
become significant.  To compute accurately $\int dr \, dL/dr$, the
data of our simulation must be both adjusted so as to correspond to
true inflow equilibrium and supplemented by an extension to larger
radius to account for the substantial radiation from radii larger than
$25M$.  Because the time-averaged fluid-frame emissivity
in the simulation tracks the NT
model so closely for $5M \leq r \leq 12M$, we define the simulation
luminosity as its $dL/dr$ integrated from the horizon to $r=12M$ plus
the NT luminosity at the mean accretion rate from $r=12M$ outward.

Given that definition, we find that the efficiency with which
this simulation generated light reaching infinity, averaged from
7000--$15000M$, was 0.151.  This number is $6\%$ greater than
the NT figure, which is 0.143 after allowing for photon capture.

\subsection{Extrapolating to the complete radiation limit}
\label{sec:extrap}

    As discussed in the Introduction, our principal goal in this
initial simulation was to achieve as close a test as possible of
the effect of the ISCO stress boundary condition on the radiative
efficiency.  We must now evaluate the degree to which our only
approximate replication of the other NT assumptions affected this test.
Time- and azimuthal-averaging should provide a good approximation
to a stationary state and axisymmetry; incomplete radiation of
the dissipated heat is our principal concern here.

    We have already seen that our radiation rate closely tracks the
NT radiation rate in the main disk body, but is about $10\%$ lower.
Thus, to extrapolate to complete radiation, the emission from this
portion of the flow should be increased by this amount.  Near the
ISCO, where the effects of the stress boundary condition become
important, we cannot use this comparison method to estimate the
magnitude of the retained heat.  Instead, we observe first that
at the ISCO the mean Thomson optical depth through the disk in our
simulation was $\simeq 500\dot m$, where $\dot m$ is the accretion
rate in Eddington units.   The corresponding diffusion time is
$\simeq 0.7\dot m$ orbits.  At the same place, the inflow rate
is $\simeq 0.6\Omega = 1.2\pi/P_{\rm orb}$.   Thus, the photon
diffusion time near the ISCO in a real disk should be shorter than the inflow
time---and shorter than our toy-model cooling time $\Omega^{-1}$---for
all accretion rates below Eddington.  A second standard of comparison
may be derived from the magnitude of the retained heat.  We found
earlier that the accretion-weighted mean specific enthalpy is $\simeq 1 + 0.018$
at $r\simeq 2M$.  That the retained heat is $\simeq 10\%$ of the binding
energy there is consistent with the fact that $\simeq 10\%$
of the heat dissipated in the main disk body is left unradiated.
Combining these two arguments, we might expect that in the limit of
truly complete radiation of dissipated heat, the efficiency
could have been greater by as much as 0.02, rising perhaps to $\simeq 0.17$,
$20\%$ above the classical number as adjusted for photon capture.

Additional heat is created in the plunging region (the mean
accreted specific enthalpy rises from 1.02 at the ISCO to $\simeq 1.03$
at the horizon), but, as we have already seen,
the fraction of photons escaping from regions so close to the horizon to
infinity is relatively small, so only a small part of the additional 0.01
in rest-mass equivalent is likely to reach distant observers.

We might also ask what effect truly radiating all the heat would have on
electromagnetic energy fluxes.  To approach this question we begin
by considering it from the point of view of the classical (NT) theory of
accretion, where much attention is paid to the $r$--$\phi$ component
of the stress tensor $T^\mu_\nu$, but little is said about other components
except for the assumption that the stress tensor is orthogonal to the four-velocity,
$u_\mu T^\mu_\nu = 0$.  As \citet{2008arXiv0801.2974B} pointed out, this assumption
is consistent with the sort of stress NT had in mind, i.e., ordinary viscosity,
but not necessarily with other physical stress mechanisms.  In particular,
it is inconsistent with MRI-driven MHD turbulence: the electromagnetic
stress tensor contains a term $||b||^2 u^\mu u_\nu$, which is manifestly
{\it not} orthogonal to the four-velocity; in addition, the turbulence
entails another (generally rather smaller) contribution to the stress tensor
$(\rho h + ||b||^2)\delta u^\mu \delta u_\nu$, where $\delta u^\mu$
is the fluctuating part of the four-velocity.  Described in
more qualitative terms, the classical theory accounts for the energy
flow due to the work done by the stress, but not the energy flow due
to the advection, by the mean flow, of an energy density associated with
the stress mechanism.

As numerous numerical studies of the MRI-driven turbulence have
shown, the fluid-frame ratio
$\alpha_{\rm mag} \equiv 2\langle b^r b_\phi \rangle/\langle ||b||^2\rangle
\simeq 0.2$--0.3 in the disk body, rising by factors of a few
in the plunging region (e.g. \cite{HK02}).  At the order of magnitude level,
the ratio of the advected magnetic energy flux to the magnetic work
is $\sim u^r/(\alpha_{\rm mag}r u^\phi)$, which is very small in the disk body, but
rises sharply near the ISCO and in the plunging region.   In this
simulation, we find that the time-averaged advected magnetic energy flux
per unit rest-mass is 0.03 at the ISCO, a significant contribution to
the energy budget.

To complete our extrapolation to complete radiation therefore means
that we need to determine how $u^r$ near the ISCO might change when that
limit is taken at fixed accretion rate.  Fixing the accretion rate means
that the vertically-integrated stress does not change, and we do not need
to estimate how the stress would change as a function of the disk's thermal
state.  If $u^r$ near the ISCO depends
primarily on the shape of the potential, the advected magnetic energy flux
per accreted rest-mass would remain roughly the same.   On the other
hand, if $u^r$ in this region depends on the gas thermal content in
the sense that it increases with increasing temperature,
more complete radiation would also lead to a smaller rate of magnetic
energy advection, and therefore to a larger net outward Poynting
flux and a larger amount of energy available for dissipation.
Which of these possibilities lies closer to the truth (and under
which circumstances) remains to be determined.

\section{Summary and Implications}

      Global disk simulations have for many years focused on dynamical
effects, i.e., angular momentum transport leading to inflow.   To link
them to observations, however, requires including considerations of
thermodynamics, for the energy to radiate photons is drawn from the thermal
energy of the gas (whether or not the particle distribution functions
are in fact near those of thermal equilibrium).  By combining an
energy conserving algorithm with an explicit cooling function in a new
simulation code, \texttt{HARM3D}, we are able to begin the first steps
toward drawing that connection.

      In this first application of our new technique,
we have found that a disk with $H/r \simeq 0.1$ accreting onto a black
hole with spin parameter $a/M = 0.9$ carries thermal and magnetic
energy past the ISCO at a rate $\simeq 0.05$ per unit rest-mass,
while producing radiation that reaches infinity at a rate $\simeq
0.15$ per unit rest-mass.  These numbers contrast with those of the
classical NT model, in which the flow carries no thermal or magnetic
energy, and for $a/M=0.9$ radiates $\simeq 0.14$ per unit rest-mass
to infinity.  Determining the observed luminous efficiency of
a more realistic accretion disk, as opposed to the ideal NT model,
depends on the careful assessment of several potentially offsetting
effects.  First, additional thermal, magnetic, and radiated energy can
be drawn from the orbital energy by magnetic stresses that can persist
through the location of the ISCO and all the way down to the horizon.
However, only a fraction of that energy need be radiated, with much of
the remainder retained as heat and magnetic field captured by the hole.
Next, even if there is enhanced photon production near and inside the
ISCO, for this particular spin, the combination of comparatively high
capture probability and gravitational redshift means that little radiation
from inside the ISCO reaches infinity.  For lower spin holes the ISCO is
further from the horizon and the plunging region can be more effectively
represented in the luminosity at infinity \citep{2008arXiv0801.2974B}.

     These results have implications for the spectral shape of the
emitted radiation.  Generically, the effect of the continuing stresses
is to move the radius of peak emission inward and raise the fluid-frame
effective temperature at that location.   For example, in this instance
the maximum in $dL/dr$ (after allowing for photon capture and all
Doppler shifts) moves from the NT prediction of $r \simeq 4.3M$ to $r
\simeq 3.5M$.  Similarly, the fluid-frame flux at the peak of $dL/dr$
is about $30\%$ greater ($7\%$ higher effective temperature) in the
simulation data than in the NT model.  In terms of the radiation edge
terminology introduced by \cite{KH02} and \cite{2008arXiv0801.2974B}, we find that
$95\%$ of the radiation reaching infinity is produced outside $r=2.75M$,
in contrast to $3.6M$ in the NT model.

     In a previous study, \cite{2008arXiv0801.2974B} used the 
stress distributions observed in an ensemble of disk 
simulations to estimate the dissipation that might be associated with
those stresses, and from this the 
accretion efficiency and maximum temperature
in the spectrum reaching infinity.  After accounting for photon
capture and Doppler-shifting effects, they found
that, depending on the particular simulation examined and the topology
of the initial magnetic field, the luminosity reaching infinity could be
anywhere from $20\%$--$100\%$ greater than NT when $a/M = 0.9$.  The low
end of this range was produced by an accretion flow whose initial field
was entirely toroidal, the high end by an accretion flow whose initial
field was a large dipolar loop, as in the present simulation.  
Thus, there is a sizable
gap between their estimate of the radiative efficiency and ours.

     Applying Beckwith et~al.'s expression to our data leads to a
prediction for the dissipation rate very similar to theirs\footnote{In
fact, the accretion rate histories of the two simulations are remarkably
similar, suggesting that the underlying physics imposes a long-term order
despite the significant difference in computational algorithms.}.  The fact
that our radiation rate is considerably less than this prediction
suggests that the simple {\it ansatz} used by Beckwith et~al. to
directly compute dissipation from stress and equate
dissipation with radiation is a simplification that likely overestimates
the net emission.  For example, because our cooling function's radiation
rate is at most comparable to the inflow rate near and inside the ISCO,
not all the heat dissipated in that region can be radiated.
However, even if all the heat generated in this simulation were radiated,
the increase in efficiency relative to NT would be only $\simeq 20\%$.
In addition, not all the work done by the stress necessarily goes into a
form that is dissipated.  Kinetic and magnetic energies can be advected with
the accretion flow into the black hole, producing no effective increase in\
overall efficiency.  This point is closely related to the issue
of advected energy discussed in \S~\ref{sec:extrap}.

Framed in the context of predictions for real accretion flows in Nature,
these questions emphasize the importance of realistic dissipation and
radiation physics for obtaining more accurate accounts of radiation
associated with accretion.  In the vicinity of the ISCO, where the
energy available for release is largest, one cannot say with confidence
that in general the dissipation and cooling times are shorter than
the inflow time.  Moreover, both processes are likely to depend on the
detailed circumstances pertaining to any particular accreting black hole,
so that there may not be a single efficiency number applicable to all
black holes of a given spin.

    In sum, we have shown that by use of a toy-model optically-thin
cooling function, it is possible both to control the thickness of the
accretion flow and to tally (approximately) the rate at which radiation
can be produced by dissipation in the flow.  At relatively large radii,
where the inflow time is long compared to the cooling time, our {\it
ansatz} of substituting gridscale dissipation for genuine microphysics
and radiating the heat so generated at an arbitrarily chosen rate is
capable of capturing the global energetics of accretion reasonably well.
However, at smaller radii (particularly near and inside the ISCO), where
the inflow time can be comparable to the cooling time, use of realistic
dissipation and radiation rates can be more important.

     Having demonstrated the technical feasibility
of this approach, we will next employ it to explore
more fully how accretion onto black holes depends on disk thickness
and on black hole rotation rate.  In this context, we point out that
although there is a standard notation for describing black hole rotation
(the spin parameter $a/M$), there are several extant definitions of
the scale-height, differing from one another by factors of order unity.
We use the vertical density moment; standardization of this
definition would be of benefit so that
different calculations can be compared quantitatively without
confusion.

    Lastly, we remark that in this paper we have set aside the
fact that photons are not the only form in which energy
can be sent to infinity from the vicinity of black holes.  Accreting
black holes are also capable of driving mass motions, often relativistic,
that can carry significant power in Poynting flux.  Simulational work
exploring the associated luminosity has already begun \citep{2004ApJ...611..977M,
2006ApJ...641..103H,2008ApJ...678.1180B}.  In future work, we will use
the new simulation code introduced here to relate the energetics of those
outflows more closely to the accretion energy budget.

\begin{acknowledgements}
This work was supported by NSF grant AST-0507455 (JHK) and PHY-0205155
(JFH).  We thank 
Charles Gammie for many enlightening discussions and an early version 
of the \texttt{HAM} code.  Exploratory simulations were performed on the
DataStar cluster at SDSC and the Woodhen cluster at Princeton University, 
while the TeraGrid T3 and Abe clusters at NCSA were used for our production runs.

\end{acknowledgements}

\bibliography{bib}

\appendix
\section{The Radiative Transfer Calculation}
\label{sec:radi-transf-calc}
Our method for calculating the radiative transfer closely follows 
the one described by \cite{2007CQGra..24..259N}.  We have made
many changes to the code, including the ability to use 3D simulation 
data, different emission models (such as the one explained here), and 
many optimizations that have made it significantly faster. 

The algorithm integrates geodesics from the observer's camera through the 
source domain---our simulation data.   These geodesics point 
toward the camera and the future.  A geodesic represents  a path along 
which a bundle of photons travel. The Lagrangian form of the 
geodesic equations is used:
\beq{
\pderiv{x^\mu}{\lambda} = N^\mu  \quad , \quad 
\pderiv{N_\mu}{\lambda} = {\Gamma^\nu}_{\mu \eta} N_\nu N^\eta    \quad , \quad 
\label{geodesic-eq-first-order}
}
where $x^\mu$ is the world-line of the photon bundle and $N^\mu$ is 
the geodesic's tangent vector parameterized by the affine parameter $\lambda$.

Since there is no absorption or scattering, the radiative transfer 
equation takes the form
\beq{
\deriv{\sI}{\lambda} = \sJ(\lambda)  \quad ,  \label{gr-rt-eq}
}
where $\sI = I_\nu / \nu^3$ is the Lorentz invariant intensity, 
$I_\nu$ is the specific intensity, $\sJ=j_\nu/\nu^2$ is the invariant emissivity, 
$j_\nu$ is the emissivity, and $\nu$ is the local frequency of the photon. 
For the purposes of calculating the bolometric luminosity, we consider
only line emission.  We can assume either constant emission frequency (e.g. 
Fe K$\alpha$ fluorescence) where we must integrate over all frequencies
at the camera, or constant observer frequency where we assume the emission
is contrived to emit at a frequency which---when redshifted to the 
camera's frame---is equal to the frequency of observation.  It is 
easy to show that both methods give the same bolometric luminosity. 
We therefore choose the latter method as it requires less computational 
effort.  

We assume that the radiation is emitted isotropically, so
$j_\nu \propto \lum/\left(4\pi\right)$ but we must also take into account the
constraint that the fluid's emission frequency is
the blueshifted frequency at the observer:
\beq{
\sJ(\lambda) = \frac{ \lum }{4 \pi \nu^2} 
      \ \delta\!\left(\nu - \nu_o/G(\lambda)\right) 
\quad , \label{line-emissivity}
}
where $G(\lambda)$, the redshift factor, is the ratio of the photon's 
energy measured by the camera to the photon's energy measured by the fluid:
\beq{
G(\lambda) = \frac{w_\mu N^\mu(\lambda_\mathrm{cam})}{ 
u_\mu(x^\mu(\lambda)) \, N^\mu(\lambda)} \quad . 
\label{redshift-factor}
}
Here, $w_\mu$ is velocity of the camera which is assumed to be static in 
flatspace;  this is a good approximation as we place the camera $10^6 M$
away from the black hole.


\end{document}